\documentclass[11pt]{article} 

\usepackage{hyperref} 
\pdfoutput=1 

\begin{document} 
\title{The Shock and the Turbulence: \\
the story of an interaction} 
\author{Iv\'an Bermejo-Moreno, Johan Larsson\footnote{Present address:
Department of Mechanical Engineering at the University of Maryland, College
Park.}\,~and Sanjiva Lele\\ 
\\\vspace{6pt} Center for Turbulence Research, \\
Stanford University, Stanford, California, 94305, USA} 
\maketitle 






The interaction of turbulence with a shock wave is a fundamental problem in
fluid mechanics relevant to a wide range of fields and applications including
aeronautics (supersonic flight and propulsion), astrophysics (supernovae
explosions, accretion shocks), nuclear physics (inertial confinement fusion),
and medicine (shock wave lithotripsy).  The canonical problem of isotropic
turbulence passing through a nominally planar shock is the simplest
configuration that isolates the fundamental physical phenomena at play in such
interaction. The controlling parameters of the interaction are the Taylor
microscale Reynolds number, $Re_{\lambda}$, and turbulent Mach number, $M_t$, 
of the incoming isotropic turbulence and the mean Mach number, $M$, that
characterizes the shock strength.

The study of this canonical problem is addressed in~\cite{Larsson_Lele_2010}
and~\cite{Larsson_etal_2013} through direct numerical simulations (DNS) of
varying $Re_{\lambda}$ (from 40 to 70), $M_t$ (from 0.05 to 0.38) and $M$ (from
1.05 to 6).  The linked
\href{http://arxiv.org/src/1310.0925v1/anc/TheShockAndTheTurbulence_HighResolution.mpg}{Video 1} (for a lower
resolution version, see
\href{http://arxiv.org/src/1310.0925v1/anc/TheShockAndTheTurbulence_LowResolution.mpg}{Video 2}), was obtained
from a simulation performed with $Re_{\lambda}=40$, $M_t=0.38$ and $M=1.5$ on a
grid containing 1040$\times$384$^2$ points, and highlights some features of this
interaction in four different sections:

\begin{itemize}
\item \textit{Shock meets turbulence}: the shock, which is nominally planar and normal to
the mean velocity by which the isotropic turbulence is advected, is corrugated
by the passage of turbulence. Local packets of faster-than-average
velocity push the shock downstream, leading to zones of stronger interaction.
The opposite holds for slower-than-average packets of the incoming fluid, for
which the shock is locally pulled upstream, leading to a locally weaker
interaction in those regions.  For the $M_t$ under consideration, holes appear
in the shock, breaking its topological structure.  Across these holes, the shock
is locally replaced by a smooth compression or by multiple weaker shocks. In
this section of the video, the time evolution of the shock is visualized in
isolation as it is corrugated by the action of the incoming turbulence. The
shock is identified by contours of high values of negative dilatation. The color
map corresponds to the local density jump across the shock, from low (in blue) 
to high (in red). The pdf of the density jump across the shock (weighted by
surface area) shows its skewed nature towards weaker events (in blue) that
dominate over the strong events (in red).

\item \textit{Turbulence meets shock}: as the turbulence passes through the shock wave,
it is also significantly altered. In particular, the turbulence kinetic energy
is amplified, although the downstream evolution of the
streamwise and transverse contributions (Reynolds stresses) differs. The former
follows a non-monotonic evolution, with a peak at a certain distance downstream
of the shock, consequence of the transfer of energy from acoustic to vortical
modes, whereas the latter peaks immediately after the shock and shows a
monotonic (viscous) decay farther downstream. Turbulent eddies are shown in
this section of the video, as educed by isocontours of the Q criterion, and
colored by the magnitude of the streamwise velocity. A scaling of Q is applied
along the streamwise coordinate to counteract the viscous decay that
dampens the turbulence significantly for the relatively low Reynolds number
attainable in these simulations. This scaling is meant to better visualize 
the turbulent eddies throughout the computational domain for a uniform
isocontour value.

\item \textit{The evolution of vortex structures}: in this section of the video two
turbulent eddies are isolated and tracked individually. One is nearly parallel
to the shock whereas the other is nearly perpendicular to it. The time evolution
of the average vorticity integrated over the surface of each eddy is shown
simultaneously. For the vortex nearly-parallel to the shock, there is a sharp
amplification of the integrated vorticity magnitude, followed by a monotonic
decay in time. On the other hand, the vortex locally normal to the shock
presents a smoother increase of the integrated vorticity, peaking at a later
time, when a significant portion of the eddy has already traversed the shock
wave.

\item \textit{The parallel lives of fluid particles}: the final section of the video
shows the trajectories of three individual fluid particles that cross the shock
at the same instant but through regions of different relative strength. The
red particle crosses through a strong interaction region; the green particle
crosses through a region of locally weak shock and the blue particle passes
through a shock hole. It is noticed that the particle going through the stronger
interaction travels faster downstream and experiences, as expected, a 
higher jump in entropy than the particle crossing through the weak interaction
region.  The particle that passes through the shock hole experiences an
isentropic compression, without any jumps. In fact, for this particle, the
entropy is first decreased across the shock hole, from a higher value relative
to the other two particles analyzed.  Farther downstream, its entropy increases
recovering the pre-shock value.  Eventually, the three particles end up with
similar entropy values. The dilatation trace in time for each particle also
confirms the strong and weak events experienced by the first two particles, and
the passing through a shock hole of the third particle (manifested in the lack
of a noticeable dilatation peak).

\end{itemize}


\par
\noindent This research was supported by the Department of Energy Scientific
Discovery through Advanced Computing (SciDAC II) program under grant
DE-FC02-06-ER25787.  The simulations were run at the Argonne Leadership
Computing Facility (ALCF) under the INCITE program as well as at the National
Energy Research Scientific Computing center (NERSC) under the ERCAP program.
The visualization cluster Certainty at Stanford University (MRI-R2 \#0960306)
was used to generate parts of this visualization.


\begin{thebibliography}{10}

\bibitem{Larsson_Lele_2010}
J.~Larsson and S.~K.~Lele.
\newblock Direct numerical simulation of canonical shock/turbulence interaction.
\newblock {\em Physics of Fluids}, 22:126101, 2010.

\bibitem{Larsson_etal_2013}
J.~Larsson, I. Bermejo-Moreno and S.~K.~Lele.
\newblock Reynolds- and Mach-number effects in canonical shock–turbulence
interaction.
\newblock {\em Journal of Fluid Mechanics}, 717, pages 293--321, 2013.

\end{thebibliography}
\end{document}